\documentclass[12pt,aps,prl,reprint,twocolumn,superscriptaddress]{revtex4-2}

\usepackage{graphicx,psfrag,amsmath,amssymb,amsfonts,bbm,latexsym,color,dcolumn,bm,mathbbol}
\usepackage[framemethod=tikz]{mdframed}
\usepackage[colorlinks=true,allcolors=blue]{hyperref}
\usepackage{soul} 
\usepackage{cancel}

\usepackage{array}
\usepackage{booktabs}
\usepackage{footnote}
\usepackage{threeparttable}

\usepackage{natbib}
\usepackage{amsmath,amssymb}             
\usepackage{color}

\newcommand{\hphicthree}{h_{\phi}^{\rm cr, 3}}
\newcommand{\ratio}{\zeta}

\usepackage{enumerate}

\definecolor{szlcolor}{rgb}{0.8, 0, 0.6}

\definecolor{bluecolor}{rgb}{0, 0, 1.0}

\definecolor{redcolor}{rgb}{1, 0, 0}

\usepackage[normalem]{ulem}
\definecolor{jfcolor}{rgb}{0.1, 0.0, 0.9}

\begin{document}

\title{Tilt-induced clustering of cell adhesion proteins}

\date{\today}

\author{Shao-Zhen Lin}
\affiliation{Aix Marseille Univ, Université de Toulon, CNRS, CPT (UMR 7332), Turing Centre for Living systems, Marseille, France}
\author{Rishita Changede}
\affiliation{Mechanobiology Institute, National University of Singapore, 117411 Singapore}
\author{Michael P. Sheetz}
\affiliation{Mechanobiology Institute, National University of Singapore, 117411 Singapore}
\affiliation{Biochemistry and Molecular Biology Department, University of Texas Medical Branch, Galveston, TX 77555}
\author{Jacques Prost}
\email{jacques.prost@curie.fr}
\affiliation{Laboratoire Physico-Chimie Curie, UMR 168, Institut Curie, PSL Research University, CNRS, Sorbonne Université, 75005 Paris, France}
\affiliation{Mechanobiology Institute, National University of Singapore, 117411 Singapore}
\author{Jean-François Rupprecht}
\email{jean-francois.rupprecht@univ-amu.fr}
\affiliation{Aix Marseille Univ, Université de Toulon, CNRS, CPT (UMR 7332), Turing Centre for Living systems, Marseille, France}

\begin{abstract} 
Cell adhesion proteins are transmembrane proteins that bind cells to their environment. These proteins typically cluster into disk-shaped or linear structures. Here we show that such clustering patterns spontaneously emerge when the protein sense the membrane deformation gradient, for example by reaching a lower-energy conformation when the membrane is tilted relative to the underlying binding substrate. Increasing the strength of the membrane gradient-sensing mechanism first yields isolated disk-shaped clusters and then long linear structures. Our theory is coherent with experimental estimates, suggesting that a tilt-induced clustering mechanism is relevant in the context of cell adhesion proteins. 
\end{abstract}

\maketitle

\textit{Introduction} -- Cell adhesion is mediated by transmembrane proteins, called cell adhesion proteins. These bind cells to one another, e.g. cadherins, or to the extracellular matrix, e.g. integrins. 
Cell adhesion proteins are known to play a central role in several processes critical in the development and maintenance of tissues and organs, such as homeostasis, immune response, and wound healing \cite{Gumbiner1996,Ladoux2012,Schwarz2013,Sun2019,Janiszewska2020}. These proteins perform these various functions by triggering cell signalling through conformational changes \cite{Changede2017}, first at the level of individual proteins -- through activation events -- but also at the collective, supramolecular level -- through the formation of large-scale structures, such as integrin-based focal adhesions and cadherin-based desmosomes \cite{Riveline2001,Maheshwari2000,Biswas2015,Bihr2015,Kechagia2019,Mamidi2018}.

In this article, we propose a generic model for the clustering of cell adhesion proteins upon interactions with a flat substrate. Our approach is in the same spirit as the pioneering work of Bruinsma and Sackmann \cite{Bruinsma2001}, but we take an explicit account of the conformation of binding proteins with respect to the membrane. Indeed, these may not strictly be orthogonal to the cell membrane; for instance, integrins have been shown to develop a mean \textit{tilt} with respect to the local normal to the membrane surface \cite{Boggon2002,Wu2010,Xu2016,Tang2018,Moore2018,Kanchanawong2023,Fenz2017}.

Considering such a tilt effect, we show that a characteristic cluster size emerges and that no ripening takes place. This contrasts with previous work on the subject \cite{Bihr2012,Bihr2015}, which considered the problem of clustering from the point of view of nucleation theory, i.e. determining the problem of reaching a critical size beyond which clusters grow. 

Our results also agree with a set of recent experiments on spreading cells \cite{Changede2015, Yu2015, Changede2019}, which show that, at the leading edge, integrins form \textit{stable}, circular, clusters with a well-defined $\approx100 \, \rm nm$ diameter. These clusters form within $3 \ \rm min$, a time short compared to their lifetime. 

Our theoretical analysis shows that the tilt effect of cell adhesion proteins yields an effectively negative surface tension that underlies the formation of stable adhesion clusters. Our numerical simulations validate our analytical bounds for cluster formation and further predict a variety of stable clustering patterns, including the experimentally observed circular patches and long linear structures, as well as curved lines, rings, Turing-like patterns, or cross-linked networks, depending on the tilt-induced negative surface tension and the chemical potential of protein-substrate binding.

\begin{figure}[t!]
\centering
\includegraphics[width=8.6cm]{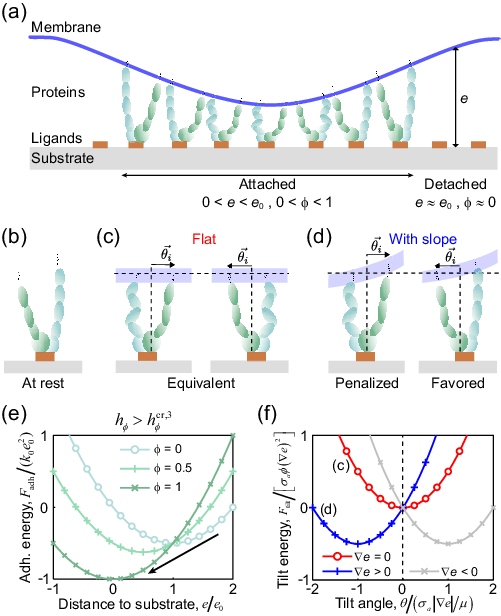}
\caption{\label{fig_ModelSketch} 
Model sketch. 
(a) Sketch of the membrane (blue), cell adhesion proteins (green and cyan), ligands (brown), and substrate (grey) model system. 
(b) Zoom on the cell adhesion protein at rest, displaying two arms (mimicking the $\alpha$ and $\beta$ arms of integrins).
(c) Zoom on two protein conformations that are equivalent in the case of a flat membrane. 
(d) Zoom on two protein conformations; in general, these are not to be expected to be equivalent in the presence of membrane height gradient; here, a positive preferred tilt corresponds to the right-hand-side conformation to be favored.
(e) The adhesion energy $F_{\rm adh}$ (see Eq. \eqref{eq:fadh}), as a function of $e$ in the case $h_{\phi} > h_{\phi}^{\rm cr,3}$. A black arrow indicates a shift from an unbound state ($\phi = 0$) to a bound state ($\phi = 1$). 
(f) The tilt free energy, $F_{\rm tilt}$ (see Eq. \eqref{eq:tiltenergydensity}), as a function of the tilt angle, $\theta$, at given values of $\nabla e$ (1D case).
}
\end{figure}

\textit{Model} -- Our model is defined in terms of two fields: $\phi \left( \bm{x} \right) \in (0,1)$, the fraction of bound cell adhesion proteins among other molecules, and $e\left( \bm{x} \right)$ measure the height of the membrane with respect to the substrate, with $e \rightarrow 0$ for a fully adhered membrane ($\phi \rightarrow 1$), and $e \rightarrow e_0$ for a fully de-adhered one ($\phi \rightarrow 0$), see Figs. \ref{fig_ModelSketch}(a,b). 

We will consider a total free energy for the membrane--protein--substrate system in the form: 
$F[e,\phi] = F_{\rm FH}[\phi] + F_{\rm Hel}[e] + F_{\rm int}[e,\phi]$, where $F_{\rm FH}$ accounts for an entropy of mixing (which depends on $\phi$ only), $F_{\rm Hel}$ for the mechanical energy of membrane deformation
(which depends on $e$ only), and $F_{\rm int}[e,\phi]$ for interactions (which will couple the $\phi$ and $e$ fields). 

More specifically, we derive the following expression for the entropy associated with protein binding:
$F_{\rm FH} = \int \mathrm{d}^2\bm{x} \{ (k_B T/a) \left[ \phi\ln\phi + (1-\phi)\ln(1-\phi) \right]+ D_{\phi}\left(\nabla\phi\right)^2/2 \}$ (see Supplemental Material \cite{SM}, Sec. I), where $k_B T$ is the thermal energy (as in Flory-Huggins theory \cite{Huggins1941,Flory1941}), $a$ is the inverse areal density of binders,
and $D_{\phi}$ is a gradient energy coefficient that controls the width of the interface of the bound proteins clusters \cite{Raote2020}. We model the membrane through the classical Helfrich free energy \cite{Helfrich1973,Weikl2018},
$F_{\rm Hel} = \int \mathrm{d}^2\bm{x} [\sigma (\nabla e)^2 / 2 + \kappa\left( \nabla^2 e -c_0 \right)^2 / 2 ]$, 
where $\sigma$ is surface tension, $\kappa$ the bending stiffness, and $c_0$ the spontaneous curvature of the membrane. 

We now consider interactions between the membrane and density fields, through molecular adhesion and molecular tilt: $F_{\rm int} = F_{\rm adh} + F_{\rm tilt}$. 

\begin{figure}[t!]
\centering
\includegraphics[width=8.6cm]{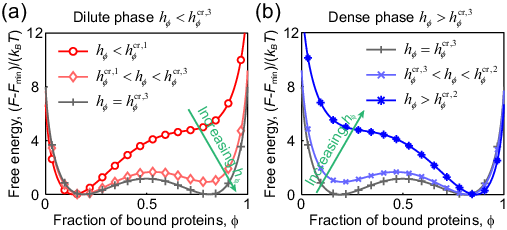}
\caption{\label{fig_FreeEnergyProfiles} 
Sketch of the free energy profiles $(F-F_{\min})/(k_B T)$ as a function of $\phi$ at different values of $h_{\phi}$, for the low temperature case ($T < T^{\rm cr}$). 
Here we assume $\phi(\bm{x}) = \phi$ and $e(\bm{x}) = e_0 (1-\phi)$; $F_{\min} = \underset{\phi \in \left( 0,1 \right)}{\mathop{\min }}\,\left\{ F\left( \phi  \right) \right\}$. 
(a) Low chemical potential case ($h_{\phi} < h_{\phi}^{\rm cr,3}$), which corresponds to a dilute phase. 
(b) High chemical potential case ($h_{\phi} > h_{\phi}^{\rm cr,3}$), which corresponds to a dense phase. 
See also Movie S1. 
For the high temperature case ($T > T^{\rm cr}$), see Fig. S3. 
}
\end{figure}

\paragraph{Adhesion} 
Following established literature, we expect the  conformation of bound proteins to be coupled to the membrane height. Adhesion molecules typically pin the membrane at a relatively short distance ($\sim 10 \ \mathrm{nm}$), as compared to the free membrane ($\sim 100 \ \mathrm{nm}$), which, in cells, interacts with the substrate due to glycocalyx steric interactions \cite{Bihr2012}. 
Building upon ref. \cite{Bihr2012}, we propose the following generic free energy for such adhesion-mediated interaction between the membrane and a flat substrate
\begin{equation}
F_{\rm adh} = \int \mathrm{d}^2\bm{x} \left\{ \frac{1}{2}k(\phi)e^2 - k(\phi) e_0(\phi) e - h(\phi) \right\}, \label{eq:fadh}  
\end{equation}
where $e_0$ stands for the membrane rest-length height (as measured from the attached height), $k$ for the membrane elastic constant, and $h$ for the chemical potential of protein-substrate binding. These parameters are expected to be functions of $\phi$. In the context of integrins, we expect $e_0(\phi)$ to be a decreasing function of $\phi$ \cite{Bihr2012}.
Here, we focus on the case of a constant $k(\phi) = k_0$, and linear relations $e_0(\phi) = e_0 (1-\phi)$ and $h(\phi) = h_\phi \phi$, see Fig. \ref{fig_ModelSketch}(e). We refer to $h_{\phi}$ as a chemical potential; increasing $h_{\phi}$ shifts the energy minima from $\phi \approx 0$ to $\phi \approx 1$ (Fig. \ref{fig_FreeEnergyProfiles} and Fig. S3; Movie S1). 

\paragraph{Tilt} 
We propose that the conformation of bound proteins is also affected by the local \textit{gradient} in the membrane height. We focus on a minimal and generic description that encompasses the onset of a local orientation order, called tilt and denoted $\vec{\theta}(\vec{x})$, and a decrease of the total free energy in the presence of a height gradient.

Our generic approach is amenable to describing bound integrins, which anchor in the membrane through two separate arms (denoted $\alpha$ and $\beta$) of unequal length, see Figs. \ref{fig_ModelSketch}(a,b). In this context, we define the deviation angle $\vec{\theta}_i$ between the overall $\beta$ arm derivation and the local membrane normal, see Figs. \ref{fig_ModelSketch}(b-d), and define the local orientation order through local averages $\vec{\theta}(\vec{x}) = \langle \vec{\theta}_i \rangle$.

In the case of a flat membrane, we expect the global tilt vector to average to zero among all adhesion proteins, based on symmetry, i.e. $\vec{\theta}(\vec{x}) = \langle \vec{\theta}_i \rangle = 0$. 

In contrast, the presence of a height gradient breaks the symmetry and allows for the emergence of an average tilt vector $\vec{\theta}(\vec{x}) \approx \vec{\theta}_{\mathrm{opt}} \neq 0$ (Fig. \ref{fig_ModelSketch}(d)). Generically, the statistics of such an average tilt vector can be described through the following free energy 
\begin{align}
F_{\rm tilt} = \int \mathrm{d}^2\bm{x} \, \phi(\bm{x}) \left[\mu \vec{\theta} \cdot \vec{\nabla} e+\frac{\nu}{2} \vec{\theta}^2 \right], 
\label{eq:tiltenergydensity}
\end{align}
where $\mu$ and $\nu$ define the average tilt angle minimizing the free energy: 
\begin{align}
\vec{\theta}_{\mathrm{opt}} = -\frac{\mu}{\nu}\vec{\nabla} e, 
\label{eq:tiltenergydensity_optimum}
\end{align}
as represented in Fig. \ref{fig_ModelSketch}(f).  Minimizing the tilt free energy density Eq. \eqref{eq:tiltenergydensity}, the tilt effect of cell adhesion proteins results in the following free energy: 
\begin{equation} \label{eq:tilt_energy}
F_{\rm tilt} = \int \mathrm{d}^2\bm{x} \left[ -\frac{1}{2}\sigma_a \phi \left( \nabla e \right)^2 \right], 
\end{equation}
where $\sigma_a = \mu^2 / \nu$ quantifies the tilt intensity. 
Equation \eqref{eq:tilt_energy} suggests that the tilt effect of cell adhesion proteins effectively contributes to a negative surface tension $\sigma_{\rm tilt} = - \sigma_a \phi < 0$. When the sum of the passive and negative surface tension cancels, we expect to see the features of Lifshitz points \cite{Hornreich1975}.

The phenomenological parameters $\mu$ and $\nu$ can be deduced from a specific microscopic model. For example, assuming that the protein tilt follows a two-state statistics $\theta_i = \pm \theta_0$, and that the conformational protein energy is $E_i = K (\theta_i + \nabla e)^2 / 2$, the Boltzmann-average protein tilt simply reads $\vec{\theta}_{\mathrm{opt}} = \left[ (K\theta_0^2)/(k_B T) \right] \vec{\nabla} e$ (see Supplemental Material \cite{SM}, Sec. II).

\begin{figure*}[t!]
\centering
\includegraphics[width=17.5cm]{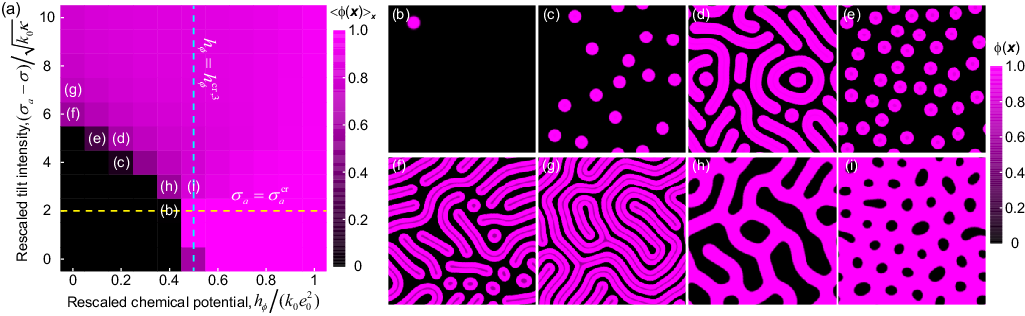}
\caption{\label{fig_PhaseDiagram} 
(a) Diagram of the fraction of cell adhesion proteins averaged over the whole space (dark, $\langle \phi \rangle = 0$; light magenta, $\langle \phi \rangle = 1$), as a function of $h_{\phi} / (k_0 e_0^2)$ (i.e. the chemical potential to membrane deformation energy cost  ratio), and $(\sigma_{a} - \sigma) / \sqrt{k_0 \kappa}$ (i.e. the tilt to attachment-induced bending surface tension ratio). 
The cyan dashed line represents the critical transition ($h_{\phi} = h_{\phi}^{\rm cr,3}$; see Eq. \eqref{eq:CriticalHphi}) from an unbound state to a bound state in the absence of the tilt effect. The yellow dashed line represents the critical transition ($\sigma_a = \sigma_a^{\rm cr}$; see Eq. \eqref{eq:CriticalSigmaA}) from a fully bound pattern to a cross-linked pattern at a high chemical potential regime, $h_{\phi} > h_{\phi}^{\rm cr,3}$. 
(b-i) Typical patterns of cell adhesion proteins clustering, where the color code corresponds to the fraction $\phi(\bm{x})$ of cell adhesion proteins at the position $\bm{x}$. Simulation domain size = $L \times L$ with $L = 16 e_0 = 1280 \ \rm nm$. 
See Table S1 in Supplemental Material \cite{SM} for parameter values. 
}
\end{figure*}

\textit{Results} -- We are interested in the minimum energy state of the system. We first consider the stability of homogeneous steady states, denoted $(\bar{e},\bar{\phi})$. Solving for the condition $\delta F/\delta \phi = 0$, we find that, the number of homogeneous states depends on the chemical potential $h_{\phi}$ and a non-dimensional parameter, 
\begin{equation}
\zeta = \frac{a k_0 e_0^2}{k_B T} , 
\end{equation}
which quantifies the ratio of the membrane-substrate repulsion energy ($\sim a k_0 e_0^2$) to the entropic energy ($\sim k_B T$). 
For a high temperature or weak binding elasticity, that is, 
$\zeta < \zeta^{\rm cr}$ with $\zeta^{\rm cr} = 4$, there exists one and only one homogeneous state, regardless of the value of $h_{\phi}$ (Supplemental Material \cite{SM}, Sec. III). 
In such a state, the fraction $\phi$ is neither close to $0$ nor $1$; indeed, the entropy dominates over the enthalpy contribution $k_0 e^2_0$, which corresponds to the energy needed to overcome the repulsive interactions when bringing an unbound membrane close to the substrate. 

For a lower temperature or a stronger binding elasticity, i.e., $\zeta > \zeta^{\rm cr}$, the situation depends on the value of $h_{\phi}$; a single homogeneous phase exists for either $h_{\phi} < h_{\phi}^{\rm cr,1}$ (corresponding to a dilute phase, i.e. with ${\phi} \approx \exp \left[ {a\left( {{h}_{\phi }}-{{k}_{0}}e_{0}^{2} \right)}/({{{k}_{B}}T}) \right] \approx 0$ and $e \approx e_0$) and for $h_{\phi} > h_{\phi}^{\rm cr,2}$ (corresponding to a dense phase, with ${\phi}  \approx 1-\exp \left[-{a{{h}_{\phi }}}/({{{k}_{B}}T}) \right] \approx 1$ and ${e} \approx 0$). 

We find that $h_{\phi}^{\rm cr, 1}$ and $h_{\phi}^{\rm cr, 2}$ depend on $\zeta$ and $k_0 e_0^2$ (the membrane deformation energy cost) with 
\begin{equation}
h_{\phi}^{\rm cr, 1} = k_0 e_0^2 g(\zeta), \quad \mathrm{and} \quad h_{\phi}^{\rm cr, 2} = k_0 e_0^2 [1 - g(\zeta)], 
\end{equation}
where $g(\zeta)$ is a dimensionless function of $\zeta$, with values between by $0$ and $1$; in the low-temperature limit $\zeta \gg \zeta^{\rm cr}$, we find that  $g \simeq \left( 1+\ln\ratio \right)/\ratio \rightarrow 0$ (see Supplemental Material \cite{SM}, Sec. III). 

For intermediate values, $h_{\phi}^{\rm cr,1} < h_{\phi} < h_{\phi}^{\rm cr,2}$, three homogeneous states exist (Fig. \ref{fig_FreeEnergyProfiles}), with a local maxima $(e_{\rm med} , \phi_{\rm med})$ and two local minima, i.e. the dilute and dense states, denoted $(e_{\max} , \phi_{\min})$ and $(e_{\min} , \phi_{\max})$, respectively.
The free energy densities of these two energy minimum states are $f\left( {{e}_{\max }};{{\phi }_{\min }} \right) \approx -{{k}_{0}}e_{0}^{2} / 2$ and $f\left( {{e}_{\min }};{{\phi }_{\max }} \right) \approx -{{h}_{\phi }}$, respectively. 
Comparison of these two free energy densities leads to the definition of a third critical chemical potential, 
\begin{equation}
\hphicthree = \frac{k_0 e_0^2}{2} , 
\label{eq:CriticalHphi}
\end{equation}
such that, for weak chemical potentials $h_{\phi} < h_{\phi}^{\rm cr, 3}$, cell adhesion proteins tend to detach from the substrate, while for strong chemical potentials $h_{\phi} > h_{\phi}^{\rm cr, 3}$, cell adhesion proteins tend to attach to the substrate. 

Such behavior follows Landau's phenomenology of phase transitions. At low temperatures $T < T^{\rm cr}$ with $T^{\rm cr}$ defined in terms of $\zeta^{\rm cr}$, i.e., $T^{\rm cr} = a k_0 e_0^2 / (k_B \zeta^{\rm cr}) = (1/4) a k_0 e_0^2 / k_B$, an increase in $h_\phi$ leads to a first-order transition from a dilute (gas-like) phase to a dense (liquid-like) state (Fig. \ref{fig_FreeEnergyProfiles}). Beyond the critical temperature $T^{\rm cr}$ the system transitions to a disordered (gas-like) state (Figs. S3 and S4; Movie S2). Such liquid/gas paradigm is mentioned for cell adhesion binders in refs. \cite{Bruinsma2001,Brochard2002}.

We next consider the dynamic stability of the homogeneous state $\left( \bar{e},\bar{\phi } \right)$, by examining the second-order variation of the total free energy, $\delta^2 F$, expressed in the Fourier space as: 
$\delta^2 F= \int \mathrm{d}^2 \boldsymbol{q} \hat{\boldsymbol{\Phi}}(\boldsymbol{q}) \cdot {\boldsymbol{J}}(\boldsymbol{q}) \cdot \boldsymbol{\Phi}(\boldsymbol{q}) / (2 \pi)^2$, 
with $\bm{q}$ being the wave vector (for details, see Supplemental Material \cite{SM}, Sec. III). 
The eigenvalues $\lambda_{\pm}(\bm{q})$ ($\lambda_{-} < \lambda_{+}$) of the Jacobian matrix ${\bm{J}}\left( \bm{q} \right)$ determine the 
dynamic stability of the homogeneous state, with ${\rm Re}(\lambda_{\pm}) > 0$ for a stable homogeneous state. 

Using this method, we find that the homogeneous dilute state ($e_{\max}$, $\phi_{\min}$) is stable for all wave vectors $\bm{q}$; 
in contrast, near the homogeneous dense state ($e_{\min}$, $\phi_{\max}$), the smaller eigenvalue reads $\lambda_{-} \simeq e_0^2 [ k_0 + \left(\sigma-\sigma_a\right) q^2 +  \kappa q^4 ]$; the minimum of such expression reads  $\min_q \{ \lambda_{-}\} = k_0 > 0$ for $\sigma_a < \sigma$, and $\min_q \{ \lambda_{-} \} = {{k}_{0}}-{{{\left( {{\sigma }_{a}} - \sigma \right)}^{2}}} / ({4\kappa })$ for $\sigma_a > \sigma$.
This leads to the following condition for instabilities to occur, 
\begin{equation}
\sigma_a^{\rm cr} = \sigma +2\sqrt{{{k}_{0}}\kappa }, \quad \mathrm{for} \quad h_\phi>h_\phi^{\mathrm{cr},3} . \label{eq:CriticalSigmaA}
\end{equation}
In the regime of $\sigma_a > \sigma_a^{\rm cr}$, the most unstable wave number is $q_{\min} = \sqrt{{({{\sigma }_{a}}-\sigma })/{(2\kappa )}}$. 
In particular, at the critical point, $\sigma_a = \sigma_a^{\rm cr}$, the most unstable wave number reads $q^{\rm cr} = (k_0 / \kappa)^{1/4}$, corresponding to the characteristic size of cluster formation, 
\begin{equation} \label{eq:lcritical}
\ell^{\rm cr} = 2 \pi \left( \frac{\kappa}{k_0} \right)^{\frac{1}{4}} .
\end{equation}

The analytical expressions of Eqs. \eqref{eq:CriticalSigmaA} and \eqref{eq:lcritical} accurately predict the transitions in patterns observed through gradient-descent simulations, see Fig. \ref{fig_PhaseDiagram}(a) and Fig. S5. We briefly sketch our numerical method (see Supplemental Material \cite{SM}, Sec. III for details). To converge to the energy minimum, we considered an annealing dynamics, whereby $\dot{\phi}=-\delta F/\delta \phi$, and $\dot{e} = -\delta F/\delta e +\eta(\bm{x}, t)$, with a noise term $\eta \left( \bm{x},t \right)$ whose amplitude 
decays with time \cite{SimulatedAnnealing}. We simulate such dynamics within the spectral method framework on a $256 \times 256$ two-dimensional lattice, under periodic boundary conditions. We point out that the total number of adhesion molecules is not conserved through such an energy minimization process. We systematically check that the final steady state does not depend on the initial states. 

In simulations with no tilt coupling ($\sigma_a = 0$), increasing the chemical potential $h_{\phi}$ leads to a first-order-like transition from the dilute to dense homogeneous phases, but does not result in stable cluster formation, as expected (Fig. \ref{fig_FreeEnergyProfiles}; Movie S1). 
For $h_\phi>h_\phi^{\mathrm{cr},3}$, upon increasing $\sigma_a$ beyond the value of $\sigma_{a}^{\rm cr}$ predicted by Eq. (\ref{eq:CriticalSigmaA}), dilute ($\phi \approx 0$) circular patches arise within the otherwise fully adherent state ($\phi \approx 1$). In contrast, for $h_\phi < \hphicthree$, dense circular patches (modeling clusters) emerge for $\sigma_a>\sigma_a^{\mathrm{cr}}$, where $\sigma_a^{\mathrm{cr}}$ now depends approximately linear on $h_{\phi}$ (Fig. \ref{fig_PhaseDiagram}). This results in the following overall expression for the critical tilt intensity
\begin{equation}
\sigma_a^{\mathrm{cr}} = \sigma+2 \sqrt{k_0 \kappa}+\omega\left(h_\phi^{\mathrm{cr}, 3}-h_\phi\right) \mathbb{1}(h_\phi<h_\phi^{\mathrm{cr},3}), 
\end{equation}
with $\omega \approx 6 \sqrt{\kappa / k_0} / e_0^2$, as obtained by numerical estimate, and $\mathbb{1}$ is the indicative function. Above the critical $\sigma_a$ value, the circular patches (dense or dilute, depending on the value of $h_\phi$) locally arrange according to a hexagonal pattern. At higher  $\sigma_a$ values, the number of circular patches decreases due to the onset of long linear structures (sometimes ring-shaped); at even larger $\sigma_a$ values, these linear structures connect into domain-size, Turing-like patterns (Fig. S6). 

The transition from hexagonally-arranged circular clusters to linear structures resembles the behavior observed in the Swift-Hohenberg equation with a quadratic term \cite{Swift-Hohenberg1977}. Detailed connections between these models are provided in the Supplemental Material \cite{SM}.

We also find that increasing the membrane tension leads to the disappearance of patterns, either in favor of the homogeneous dilute and detached state, for $h_{\phi} < h_{\phi}^{\rm cr,3}$, or to the homogeneous dense and adhered state for $h_{\phi} > h_{\phi}^{\rm cr,3}$, see Fig. \ref{fig_PhaseDiagram} and Fig. S7; such a transition is echoed by experimental observations \cite{Delanoe2004}.

\textit{Experimental relevance} -- We consider a typical distance between binders $d = 10 \ \rm nm$ ($a = 100 \ \rm nm^2$) \cite{Changede2015}; a typical binding energy $k_0 e_0^2 a \sim 10 \ k_B T$ \cite{Bihr2012,Bihr2015}; a typical height difference $e_0 = 80 \ \rm nm$. With these values, we find that the ratio of the adhesion energy to the entropic energy $k_0 e_0^2/(k_B T/a) \sim 10$, hence satisfying $\zeta > \zeta^{\rm cr}$ (i.e. low temperature case $T < T^{\rm cr}$). These estimates also fix the substrate binding stiffness at $k_0 \sim 10 k_B T / (e_0^2 a) \sim 10^{-5} \ k_B T \cdot \rm nm^{-4}$; the effective stiffness, $k_0 a \sim 10^{-3} \ k_B T \cdot \rm nm^{-2}$, is consistent with one provided in ref. \cite{Bihr2012} (the parameter $\lambda$ therein). Further, we consider $D_{\phi} = k_B T$.

Considering a typical cell membrane tension $\sigma = 2 \times 10^{-5} \ {\rm J \cdot m^{-2}} \approx 0.005 \ k_B T \cdot \rm nm^{-2}$ \cite{Kozlov2015,Raote2020} and membrane bending rigidity $\kappa = 10 \ k_B T$ \cite{Raote2020,Steinkuhler2019,Weikl2018}, we find that adhesion is the dominant contribution to the critical tilt strength, with $\sigma^{\rm cr}_a \approx 7 \sigma \approx 0.03 \ k_B T \cdot \rm nm^{-2}$. 

Such critical tilt energy is accessible. Indeed, a $1 \ k_B T$ gain in conformational energy due the tilt then corresponds to $\sigma_a \approx 0.1 \ k_B T \cdot \mathrm{nm}^{-2} \approx 3 \sigma^{\rm cr}_a$, with a height gradient of order $1$ (for integrins, estimations for the membrane height are in the range $100 \ \mathrm{nm}$ in the de-adhered state, $30 \ \mathrm{nm}$ in the adhered state \cite{Bihr2012}, while a typical cluster diameter is $\ell \sim 100 \ \mathrm{nm}$ \cite{Changede2015}.

The estimate of the cluster length discussed in Eq. (\ref{eq:lcritical}) corresponds to $\ell^{\rm cr} \approx 160 \ \rm nm$; by construction, such value is an upper bound estimate. In our simulations, we rather observe clusters that are $\ell^{\rm sim} \approx 100 \ \rm nm$ in diameter, as observed in experiments \cite{Changede2015}.

\textit{Robustness and generalization} -- We checked that our results are robust to various changes in parameters. 
Similar results are obtained for a $\phi$-dependent elastic constant $k(\phi)$ in Eq. \eqref{eq:fadh} (see Supplemental Material \cite{SM}, Sec. IV; Fig. S8). Our framework could also be rephrased to describe the case of identical and opposite membrane deformations, hence addressing the case of cadherin binding opposing membranes. In addition, one could consider the effect of transient changes in integrin conformation occurring during the activation and deactivation processes \cite{Limozin2019}. Since activation mediates a larger spacing between the two transmembrane domains of integrins\cite{Limozin2019}, we expect activation events to modulate the membrane-gradient sensing term $\sigma_a$. We leave these aspects to further studies. 

\textit{Conclusion} -- Recent development in polarization imagery techniques allows us to infer the conformation of proteins within biological membranes \cite{Curcio2020}. Anticipating the application of such a technique to cell adhesion complexes, here we proposed a new paradigm for the relationship between molecular structure and the formation of supramolecular assemblies.

\textit{Acknowledgements} -- 
J.-F. R. is hosted at the Laboratoire Adhésion Inflammation (LAI). J.-F. R. thanks fruitful discussions with Pierre-Henri Puech. The project leading to this publication has received funding from France 2030, the French Government program managed by the French National Research Agency (ANR-16-CONV-0001) and from Excellence Initiative of Aix-Marseille University - A*MIDEX. J.-F. R. is also funded by ANR-20-CE30-0023 COVFEFE.

\end{document}